\documentclass[showpacs,preprintnumbers,amsmath,prl,
graphicx,amssymb,superscriptaddress,twocolumn,nofootinbib]{revtex4}
\usepackage{graphicx}
\usepackage{dcolumn}
\usepackage{bm}

\renewcommand{\today}{\number\day\space\ifcase\month\or January\or 
 February\or March\or April\or May\or June\or July\or August\or 
 September\or October\or November\or December\fi\space\number\year}

\begin{document}

\title{A Maximum Likelihood Analysis of Low-Energy {\sc CDMS} Data}

%%%%%%%%%%%%

\newcommand{\efi}{Enrico Fermi Institute, Kavli Institute for 
Cosmological Physics and Department of Physics,
University of Chicago, Chicago, IL 60637}
%%%%%%%%%%%%

\affiliation{\efi}
														
\author{J.I.~Collar and N.E.~Fields}\affiliation{\efi}

\begin{abstract}

An unbinned maximum likelihood analysis of  {\sc CDMS} low-energy data reveals a strong preference ($5.7\sigma$ C.L.\!) for a model containing an exponential excess of events in the nuclear recoil band, when compared to the null hypothesis. We comment on the possible origin of such an excess, establishing a comparison with anomalies in other dark matter experiments. A recent annual modulation search in {\sc CDMS}  data is shown to be insufficiently sensitive to test a dark matter origin for this excess. 
\end{abstract}

%% ENTER PACS NEXT
\pacs{95.35.+d, 85.30.-z}
%\pacs{85.30.-z, 95.35.+d, 95.55.Vj, 14.80.Mz}

\maketitle

%%%%%%%%%%%%%%%%%%%%%%%%%%%%%%%%%%%%%%%%%%%%%%%%%%%%%%%%%%%%%%%%%%%%%%

The {\sc CDMS} collaboration has recently made public a negative search for an annual modulation in low-energy signals from their cryogenic germanium detectors \cite{cdmsmod}. This effect is expected from Weakly Interacting Massive Particle (WIMP) interactions with dark matter detector targets \cite{annmod}. Observation of this WIMP signature has been claimed by the {\sc DAMA} collaboration with high statistical significance \cite{DAMA}, using low-background NaI(Tl) scintillators. The {\sc CoGeNT} collaboration recently released fifteen months of data  from underground germanium detector operation \cite{cogent3}. These display a compatible modulation \cite{cogent3,nealdan1,nealdan2}, albeit with the smaller statistical significance that would be expected from a short exposure. 

Fig.\ 6 in \cite{cdmsmod} shows, for the first time, detailed information from all eight {\sc CDMS}  germanium detectors employed in the modulation search and a previous low-energy analysis \cite{prevcdms}. Specifically, it contains the distribution of single-interaction events in the ionization energy ($E_{i}$) vs.\ recoil energy ($E_{r}$) plane that can be used to identify their origin in nuclear recoils (NR) like those expected from WIMP and neutron interactions, or electron recoils (ER) like those induced by gamma backgrounds. 

A formal assessment of the possibility that a significant WIMP-like low energy NR population might be present in {\sc CDMS} data is of particular interest. The  {\sc CDMS}  collaboration has not made public a dedicated study of this type, but has put forward arguments \cite{prevcdms} indicating that the majority of their low-energy events would originate in unrelated backgrounds. These arguments have been criticized in \cite{mycrit}. Additional objections will be presented here, arising from the analysis described below.

\begin{figure}
\includegraphics[width=8.5cm]{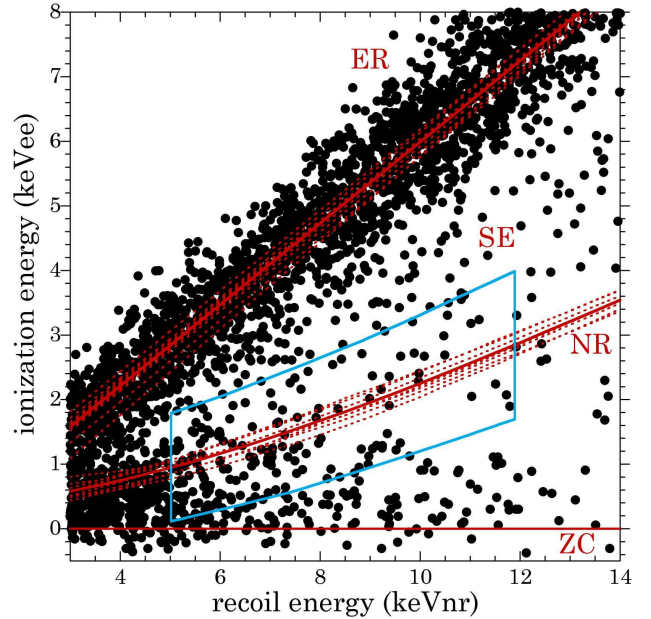}
\caption{Scatter plot of single-interaction events in all eight {\sc CDMS} detectors, digitized from individual plots in \protect\cite{cdmsmod}, using {\sc g3data} software \protect\cite{g3data}. Thick solid lines represent the approximate centroid of the electron recoil (ER) and nuclear recoil (NR) response bands, averaged over all detectors, with thin dotted lines corresponding to each individual device, averaged over the small temporal variations observed during a two-year span (Fig.\ 6 in \protect\cite{cdmsmod}). A third thick line indicates the nominal centroid for zero-charge (ZC) events, defined by the condition $E_{i}\!=\!0$ (see text for a discussion on deviations). A disproportionate fraction of events in the region $E_{i}\!<\!1.5$ keVee, $E_{r}\!>\!5$ keVnr, come from T3Z5, a detector with a markedly degraded energy resolution. The blue boundary is the (detector-averaged) region selected by {\sc CDMS} for an annual modulation search \protect\cite{cdmsmod}.}
\end{figure}

Fig.\ 1 displays the sum of these data in $E_{i}$ vs.\ $E_{r}$ representation for dark matter search runs collected over a two year span. The response across detectors to gamma and neutron calibration sources varies from device to device, as denoted by the individual centroids of the ER and NR bands (thin dotted lines). While these  variations have the effect of somewhat blurring the distinction between background sources, they are sufficiently small not to preclude an attempt to analyze summed data, in an effort to maximize the available statistics. Besides the mentioned ER and NR components, two additional sources of background are described in \cite{prevcdms}: a diffuse surface event component (SE, Fig.\ 1) due to gamma and beta interactions on detector surfaces, leading to partial charge collection, and a population of zero-charge (ZC, Fig.\ 1) events taking place on the edges of the detectors, resulting in complete charge loss due to local electric field distortions.

We employ the {\sc RooFit} library \cite{roofit}, part of the {\sc ROOT} software distribution, to perform an unbinned maximum likelihood analysis on {\sc CDMS} data. The purpose is to compare the null hypothesis (a model containing ER, SE and ZC components only) with the alternative hypothesis, which includes an additional NR component. The ER, NR and ZC components are defined similarly by probability density functions (PDFs) of the form $A(E_{r})\times G(E_{i},E_{r})$, where $A(E_{r})=A_{1}\cdot e^{-A_{2}\cdot E_{r}}$, $A_{1},A_{2}$ are positive-defined free parameters, and $G(E_{i},E_{r})=e^{-(E_{i}-C(E_{r}))^{2}\!/ 2 \sigma^{2}\!(E_{r})}$. The addition of a constant term to $A(E_{r})$  is seen to be unnecessary. The function $C(E_{r})$ returns the ionization energy of the centroid for the ER, ZC and NR bands (Fig.\ 1). This function is linear for ER and ZC ($C(E_{r})\!\!\sim\!0$ for ZC, see discussion below), and quadratic for NR. The function $\sigma^{2}\!(E_{r})=S^{2}_{1}+S_{2}\!\cdot\!C(E_{r})$, with $S_{1},S_{2}$ positive-defined free parameters, accounts for the energy resolution of the detectors, dominated by the significantly noisier ionization channel (a factor $\sim$3 larger dispersion than along the phonon channel \cite{resol,fe}). The comparatively small smearing of the data from the resolution along the phonon (recoil energy) channel is absorbed by the model through $A(E_{r})$. The functional form of $\sigma^{2}\!(E_{r})$ arises from a conventional description \cite{noise}, where $S_{1}$ encapsulates all sources of electronic noise and $S_{2}$ accounts for the statistics of information carriers (electron-hole pairs here). While ER, ZC and NR components of the model are allowed independent values of $A_{1}$ and $A_{2}$ (i.e., are allowed to grow exponentially towards diminishing $E_{r}$ independently of each other), the energy resolution parameters $S_{1},S_{2}$ are common to all components: in other words, the energy resolution is blind to the nature of an event generating a pulse of a given $E_{i}$.  Finally, based on the information in \cite{cdmsmod, prevcdms, add}, we expect the combined triggering and software cut efficiency to be sufficiently close to 100\% over the range of Fig.\ 1 to be conservatively neglected.

Following the description provided in \cite{prevcdms},  SE events are combined with those in ER through a "Crystal Ball" PDF \cite{crystal}, conveniently pre-defined within {\sc RooFit}. This approach (a Gaussian core as above to account for ER, with a power-law low-$E_{i}$ tail for SE) describes the noticeable "bleeding" of ER events into the SE population due to partial charge collection at surfaces. The distribution of ER events in energy is essentially flat for the studied range, with a peak of negligible intensity at $E_{i}\!=\!6.5$ keVee from cosmogenic activation of long-lived $^{55}$Fe \cite{cogent3,fe} as the only discernible feature.

\begin{figure}
\includegraphics[width=8.5cm]{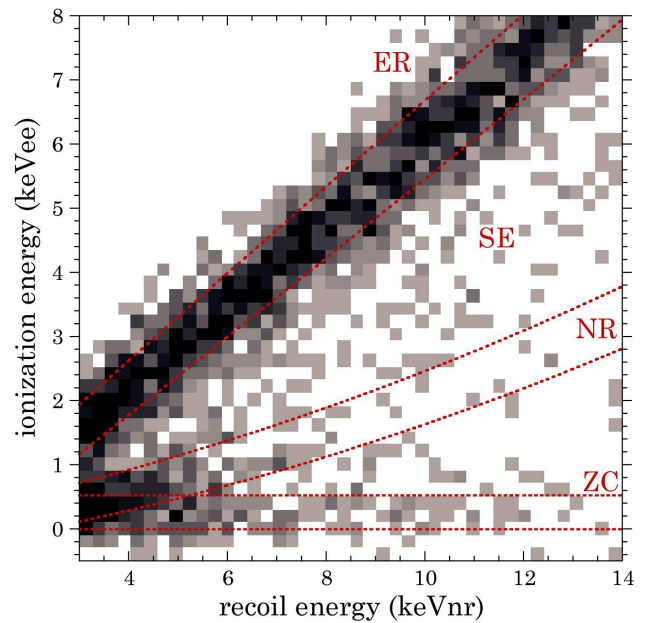}
\caption{Grayscale intensity plot equivalent to Fig.\ 1, with detector T3Z5 removed. Dotted lines represent the best-fit ($\pm1\sigma$) boundaries for the ER, NR and ZC bands. The best-fit to the electronic noise of the detectors compares well with a detector-averaged value in \protect\cite{fe} ($303\pm 30$ eV  vs.\ 293 eV, respectively). We employ the same model description as {\sc CDMS} to account for ZC events (see text).}
\end{figure}

The description above is used for the analysis of individual detectors, but an additional allowance must be made for an adequate treatment of summed detector data: while $C(E_{r})$ in principle contains no free parameters, the coefficients defining these linear or quadratic functions are allowed a modest Gaussian smear defined by the variance in their values across detectors. This accounts for detector differences in energy scale calibration, energy resolution, exposure, ER background rate, and the necessary averaging of the small temporal variations observed in their response \cite{cdmsmod}. We apply the required penalty to the goodness-of-fit when enabling this constrained fit. 

The dispersion along $E_{i}$ for ZC events in any given detector is defined by the electronic noise of the ionization channel, and as such should be symmetric around $E_{i}\!=\!0$. The effective ZC centroid can nevertheless acquire a non-zero value due to, for instance, a less-than-perfect energy calibration along $E_{i}$. We conservatively allow for a positive ZC centroid (Fig.\ 2) even when analyzing individual detectors, obtaining excellent agreement with {\sc CDMS}'s identical modeling of ZC events. Specifically, matching within $70$ eV both positive ($E_{i}\!=\!+0.7$ keVee) and negative ($E_{i}\!=\!-0.26$ keVee) $2\sigma$ ZC  boundaries for detector T1Z5, the only one for which this information has been made available \cite{presentations}. In spite of this possible small ZC shift, the symmetry of positive and negative fluctuations in detector noise about its baseline is to be expected, and well-known to workers in radiation detection \cite{staham}. In the case of {\sc CDMS} detectors it corresponds to a symmetric "zero strobe" peak \cite{zero} triggered by a ZC event registered along the phonon (recoil energy) channel only. Keeping in mind the physical origin of ZC events (low-energy electrons traveling through edge regions) we expect, if anything, less dispersion in $E_{i}$ about zero the lower the energy (shorter the electron range) of the interaction. High-energy ZC events of up to 100 keVnr remain tightly clustered about $E_{i}\!\sim\!0$ \cite{prevcdms}. The absence of significant ZC band distortion or broadening with decreasing $E_{r}$ is evidenced in {\sc CDMS} $^{133}$Ba calibrations \cite{presentations}.

It is worth mentioning at this point that the ionization yield vs.\ recoil energy plot for detector T1Z5 offered in most recent {\sc CDMS} public presentations \cite{presentations} displays markedly different boundaries for the ZC band ($\pm 2\sigma$) and NR band  (+1.25$\sigma$,-0.5$\sigma$). This, together with the chosen $E_{i}/E_{r}$ vs. $E_{r}$ representation, communicates the perception that most low-energy events below $\sim\!5$ keVnr belong to the ZC population. A more objective representation like that of Fig.\ 2 results in the immediate impression, confirmed by our formal analysis, that a large fraction of these events belongs to NR.

Table I contains the results of comparing null and alternative hypotheses via the standard log-likelihood ratio test statistic, for summed data and individual detectors. $\Lambda$ is the ratio between the likelihood of the null and alternative hypotheses, and $\Delta$d.o.f. represents the difference in degrees of freedom. The small p-values obtained from this statistic \cite{wilks} indicate that the model containing a NR component is strongly favored, a statement that can be maintained with a $>5\sigma$ C.L. for summed data.  The best-fit to the integrated NR event rate and NR exponential decay constant $A^{NR}_{2}$ seems compatible across all cases studied, once the uncertainties and detector differences (exposure, resolution) are taken into account. Fig.\ 3 displays the projections of the data and best-fit model components along the $E_{i}$ and $E_{r}$ axes.

\begin{widetext}

\begin{table}[ht]
\let\thempfootnote\thefootnote
\centering
\begin{tabular}{ | c | c | c | c | c | c | c |}
\hline
\ Detector\ & Exposure (kg-day) & \ -2 ln$(\Lambda)/\Delta$d.o.f. &\ p-value & \ $S_{1}$(eVee) & \ NR rate $>3$ keVnr (c/kg-day) & \ $A^{NR}_{2}$ (keVnr$^{-1}$) \\
 \hline \hline
 all  & 240.4 & 37.8 / 5 & $4.1\cdot 10^{-7}$ & 303 & $0.93\pm0.17$ & $0.650\pm0.081$ \\
 all$-$T3Z5  & 214.6 & 45.9 / 5 & $9.5\cdot 10^{-9}$ & 266 & $0.95\pm0.14$ &  $0.701\pm0.079$ \\
T1Z2  & 43.4 & 13.5 / 2 & $1.1\cdot 10^{-3}$ & 220 & $0.77\pm0.20$ & $0.569\pm0.139$ \\
T1Z5   & 35.0 & 12.5 / 2 & $1.9\cdot 10^{-3}$ & 201 & $0.50\pm0.17$ &  $0.714\pm0.299$ \\
T2Z3 & 28.0 & 5.7 / 2 & $5.7\cdot 10^{-2}$ & 246 & $1.31\pm0.68$ & $0.659\pm0.222$ \\
T2Z5 & 34.7 & 2.7 / 2 & $2.6\cdot 10^{-1}$ & 442 & $0.76\pm0.41$ & $0.745\pm0.355$ \\
T3Z2  & 7.8 & 3.9 / 2 & $1.4\cdot 10^{-1}$ & 333 & $3.38\pm1.24$ &  $0.705\pm0.215$ \\
T3Z4  & 29.6 & 12.8 / 2 & $1.7\cdot 10^{-3}$ & 142 & $0.39\pm0.15$ & $0.636\pm0.202$ \\
T3Z5 & 25.8 & 0.22 / 2 & $8.9\cdot 10^{-1}$ & 406 & $0.15\pm0.38$ & $2.01\pm2.44$ \\
T3Z6 & 36.1 & 3.27 / 2 & $1.9\cdot 10^{-1}$ & 228 & $0.61\pm0.31$ & $0.707\pm0.361$ \\
\hline

\end{tabular}
\caption{Comparison of null and alternative models via the log-likelihood ratio (see text). As expected, the p-value for individual detectors correlates to their electronic noise ($S_{1}$), responsible for loss of energy resolution, and with it the ability to distinguish between background components. }

\label{tab:data_summary}

\end{table}

\end{widetext}

The physical origin of an exponential excess along the NR band, concentrated below a few keVnr only, is a daunting question. A few remarks can nevertheless be made with some degree of certainty. {\sc CDMS} has recently stated \cite{prevcdms} that Monte Carlo simulations produce a negligible $<1$ event background for the dataset examined here. We agree with this statement: based on {\sc CoGeNT} simulations, scattering of underground neutrons from cavern and ($\mu$,n) origins is expected to produce a much less steep rise in NR rate towards low $E_{r}$, and a low-energy NR rate negligible when compared to the magnitude of this excess \cite{marek}. In earlier searches \cite{cdmsold}, the {\sc CDMS} collaboration found no evidence for NR events above 15 keVnr (an aggressive SE cut is only possible above $\sim10$ keVnr \cite{cdmsmod,prevcdms}), confirming that a neutron origin is highly implausible. Separately, events registering in multiple detectors ("multiples", Fig.\ 6 in \cite{cdmsmod}) are seen to spill into the low-energy NR bands. A fraction of these originate from surface events involving simultaneous beta and gamma emission (e.g., from $^{210}$Pb deposition \cite{theses}), where a low-energy surface beta can have only part of the ionization it generates collected. The subset of these episodes where the gamma escapes interaction with another detector could then give rise to a family of single-interaction events within the low-energy NR band. However, multiples within the $\pm 2\sigma$ NR detector bands and $E_{r}\!<\!6$ keVnr exhibit a flat distribution in ionization energy, as opposed to the singles distribution, which is markedly peaked around the NR centroid ($\sim\!0.6$ keVee) and well-separated from the ER band. In addition to this, when the rates of singles and multiples for each detector are examined in this same region, they exhibit a near-perfect lack of linear dependence (Pearson's correlation coefficient $r\!=\!0.08$), also indicating that the possibility of a common physical origin is remote.

The more exotic alternative of a WIMP origin can be assessed by comparison to recent anomalies in other searches. Fig.\ 4 overlaps the best-fit to the NR band component from the analysis of all summed detectors onto {\sc CoGeNT} data \cite{cogent3}. The conversion to ionization energy is done as in \cite{cdmsmod}, by using the more reliable germanium quenching factor measured by {\sc CoGeNT} \cite{dong}. Assuming this exponential distribution is the approximate response to a WIMP, we generate a {\sc CDMS} region of interest (ROI) in WIMP coupling vs.\ WIMP mass (Fig.\ 4, inset) that includes present uncertainties.

\begin{figure}
\includegraphics[width=8.5cm]{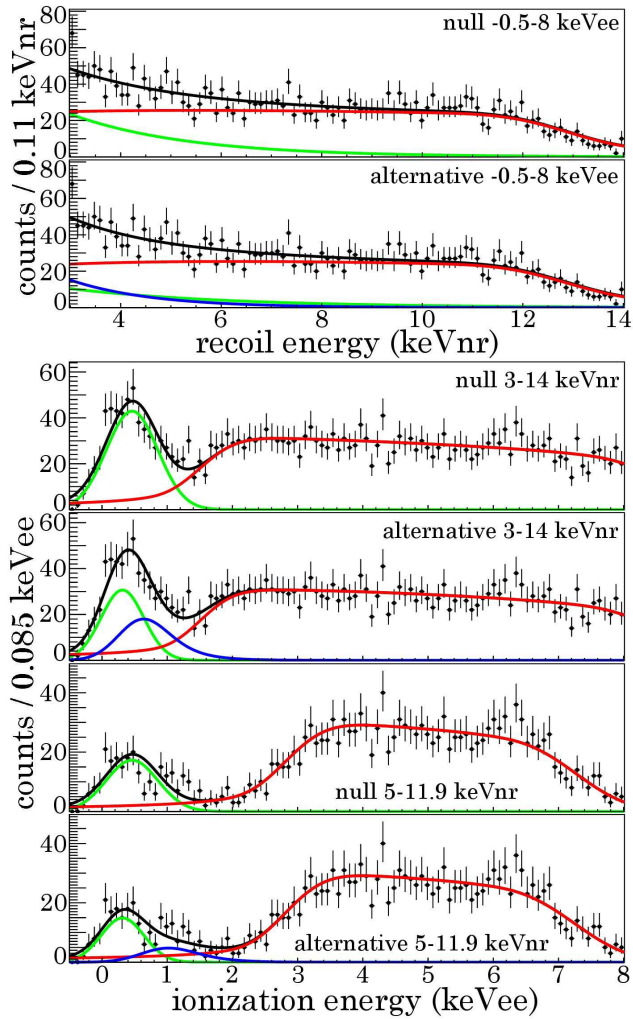}
\caption{Best-fit components of the null and alternative models, overlapped on summed data, projected on $E_{r}$ and $E_{i}$. Red: ER and SE combined through the Crystal Ball PDF \protect\cite{crystal}. Green: ZC. Blue: NR. Black: sum of components. The null model requires a large  deviation of the ZC centroid to $E_{i}\!\sim\!0.5$ keVee, hard to reconcile with adequate $E_{i}$ calibrations \protect\cite{fe} and with the mean $E_{i}$ of ZC events above $\sim 5$ keVnr, a region where their true centroid can be assessed (Fig.\ 2). The separation between ZC and NR populations is noticeable for data in the 5-11.9 keVnr analysis region used in \protect\cite{cdmsmod}.}
\end{figure}

\begin{figure}
\includegraphics[width=8.5cm]{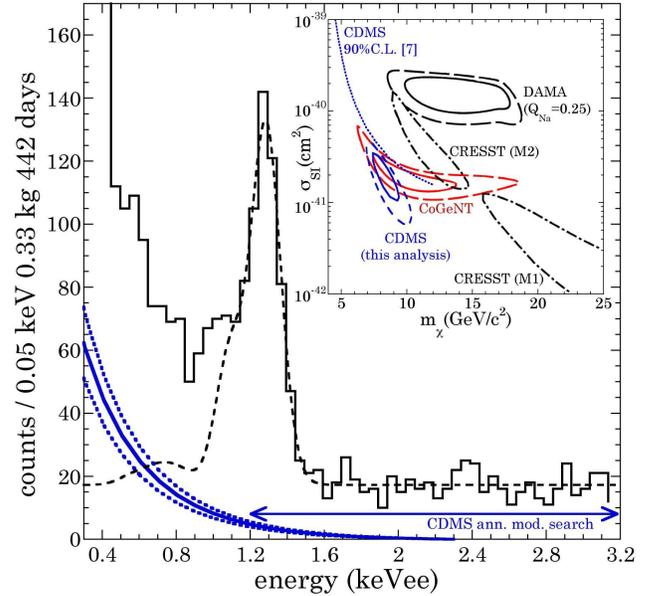}
\caption{Blue: best-fit NR component for {\sc CDMS} summed detector data, translated to ionization scale and overlapped on histogrammed {\sc CoGeNT} data \protect\cite{cogent3} after normalization to the vertical scale. Neither is corrected for efficiency next to threshold. Dotted blue lines represent the $1\sigma$ uncertainty in the parameter $A_{1}$ for NR. A dashed black line represents known {\sc CoGeNT} backgrounds (flat+cosmogenic \protect\cite{cogent3}), which provide an adequate fit to the data down to $\sim 1.2$ keVee, the lower boundary of the {\sc CDMS} annual modulation search region. Inset: 90\% and 99\% C.L. {\sc CDMS} ROI in WIMP coupling vs.\ mass (see text), including all present uncertainties except for those related to {\sc CDMS}'s energy scales \protect\cite{dong}. ROI's for {\sc CoGeNT} \protect\cite{cogent3}, {\sc CRESST} \protect\cite{cresst} and {\sc DAMA }\protect\cite{DAMA} are from \protect\cite{nealdan2}, and include the effect of a residual surface event contamination in {\sc CoGeNT} described in \protect\cite{cogent3}. The {\sc DAMA} ROI assumes a Maxwellian dark matter halo: deviations from it can displace it to lower WIMP couplings \protect\cite{nealdan1,nealdan2,spergel,mytalk}. Additional uncertainties for {\sc DAMA} exist \protect\cite{mytalk}.}
\end{figure}

Our analysis allows for a straightforward estimate of the sensitivity of the search for an annual modulation in \cite{cdmsmod}, by integrating best-fit signal (NR) and background (ER, SE, ZC) components inside {\sc CDMS}'s $\pm 2\sigma$ NR, 5-11.9 keVnr, "signal box" (Fig.\ 1, blue enclosure). We find that out of $\sim\!167$ events within, only 35\% would correspond to the putative WIMP (NR) signal. This translates into 0.035 NR events / keVnr kg day, whereas the 99\% exclusion claimed in \cite{cdmsmod} is for modulations larger than 0.06  events / keVnr kg day. In other words, even at a modulation amplitude of 100\%, the search in \cite{cdmsmod} would fail to exclude a WIMP origin for the NR excess seemingly present in {\sc CDMS} data. Important additional concerns about the search in \cite{cdmsmod} can be listed. For instance, the addition of non-overlapping time periods \cite{periods} from detectors spanning an order of magnitude in background rate within the signal box \cite{add,presentations}, the negligible overlap with the {\sc CoGeNT} spectral region containing a clear  excess of events (Fig.\ 4), or unresolved issues related to {\sc CDMS}'s energy scales \cite{dong}. However, we emphasize that the choice of signal box  boundaries, one that results in a poor signal-to-background ratio, is already sufficient to cripple its sensitivity.

In conclusion, we find that recently released 2007-2008 {\sc CDMS} data strongly support ($5.7 \sigma$ C.L.) the presence of a family of low-energy events concentrated in the nuclear recoil band. An origin in neutron scattering is highly unlikely. Their rate and spectral shape provides a match to the majority of low-energy events unaccounted for by {\sc CoGeNT} (Fig.\ 4). Both searches employ the same target material, germanium, but perform orthogonal background cuts \cite{mycrit}, enhancing the possible meaning of this coincidence. If this excess is interpreted as a WIMP signal, it falls in close proximity to other anomalies reported by recent dark matter experiments (Fig.\ 4). The favored region of WIMP mass is also of present interest in indirect searches for dark matter \cite{indirect}. We also determine that the recent search for an annual modulation signal by the {\sc CDMS} collaboration is insufficiently sensitive to exclude a dark matter origin for this excess, due to an inadequate selection of analysis region. Unsupported quantitative statements made in \cite{prevcdms,resol} about background composition in {\sc CDMS} detectors are not compatible with our findings. 

We encourage the {\sc CDMS} collaboration to continue refining their understanding of detector backgrounds and energy scales. We are indebted to M. Bellis, D. Hooper, C. Kelso, D. Moore, A.E. Robinson and N. Weiner for useful discussions. N.F. is supported by the NNSA Stewardship Science Graduate Fellowship program under grant number DE-FC52-08NA28752.


\begin{thebibliography}{9}
\bibitem{periods}Data quality cuts reduce the usable live-time of {\sc CDMS} detectors to less than 50\% of the already discontinuous detector-specific live periods shown in Fig.\ 5 of \cite{cdmsmod}. In the extreme case of T3Z2, just 10\% of the exposure implied by that figure was used for the analysis in \cite{cdmsmod}. Operational stability of detectors is an important prerequisite for a modulation search \cite{cogent3}.
\end{thebibliography}
\end{document}